\documentstyle[prl,aps,multicol]{revtex}
\input psfig
\begin{document}

\draft \preprint{G.\ Schmidt}

\title{Large magnetoresistance effect due to spin-injection into a non-magnetic
semiconductor}

\author{G. Schmidt, G. Richter, P. Grabs, C. Gould, D. Ferrand, L. W. Molenkamp}

\address{Physikalisches Institut, Universit\"at W\"urzburg, Am
Hubland, 97074 W\"urzburg, Germany}

%\author{} %\address{}

\date{\today}

\maketitle

\begin{abstract}

A novel magnetoresistance effect, due to the injection of a
spin-polarized electron current from a dilute magnetic into a
non-magnetic semiconductor, is presented. The effect results from
the suppression of a spin channel in the non-magnetic
semiconductor and can theoretically yield a positive
magnetoresistance of 100\%, when the spin flip length in the
non-magnetic semiconductor is sufficiently large. Experimentally,
our devices exhibit up to 25\% magnetoresistance.

\end{abstract}


\begin{multicols}{2}
%\section{Introduction}
Semiconductor spintronics has gained a strong boost from the recent experimental %%@
demonstration of electrical spin injection into a non-magnetic
semiconductor (NMS), using Dilute Magnetic Semiconductors (DMS) as
spin-injecting contacts\cite{injector,injector2}. However,  the
practical implications of these achievements for utilizing spin
injection in semiconductor circuits are still limited, since in
both experiments the spin polarization of the current was detected
via the circular polarization of the electroluminescence of a
semiconductor light emitting diode, and no appreciable effect of
the spin polarization on the resistance of the device could be
observed. Evidently, such an effect would be extremely useful for
the implementation of spin injection in  semiconductor transport
devices for memory and logic applications. An obvious candidate
for  implementing a spin-dependent resistance in a semiconductor
device is based on utilizing the Giant Magneto Resistance (GMR)
effect, which is well known from all-metal ferromagnetic/non-
magnetic multilayer devices\cite{gmr}. However, the practical
realization of a semiconducting GMR device has proven to be
difficult, mainly because the effect relies on utilizing
ferromagnetic contacts. We now know\cite{gs} that spin-injection
into semiconductors can only be achieved from a contact that has a
similarly low conductance as the non-magnetic semiconductor, and a
close to 100 \% spin-polarization. This excludes using
ferromagnetic metals like Fe, Co, or Ni as contact materials. As
shown in \cite{injector}, II-VI-DMSs do fulfill the 100\%
polarization requirement \cite{gs} and provide a solid means for
generating a strongly spin-polarized current in a NMS.

We have now found that the tunable Zeeman-splitting in these
paramagnetic DMSs allows for the realization of a novel
magnetoresistance effect. The effect (a large positive
magnetoresistance) is caused by the suppression of one spin
channel in the NMS. In this paper we describe the observation of
the novel effect.

\begin{figure}
\centerline {\psfig{file=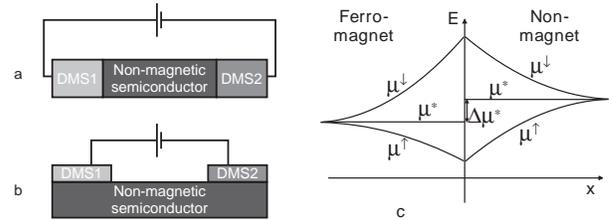,width=8.0cm}} \caption{Idealized one-dimensional structure
consisting of a non magnetic semiconductor with two attached DMS contacts. (b) Spin injection
device used in the experiment consisting of a non-magnetic semiconductor layer with two DMS
top-contacts. (c) Electrochemical potentials at a ferromagnet non magnet interface crossed by a
spin polarized current. For clarity, the linear part of the potentials was removed.} \label{fig1}
\end{figure}

Consider a device where a NMS layer is fitted with two paramagnetic DMS-contacts which can be
either non-magnetized  or magnetized in parallel by a suitable external field (Fig. \ref{fig1}a).
In such a device, the current will be either unpolarized (non-magnetized contacts) or
spin-polarized (magnetized contacts). In an NMS, electrons with spin-up and spin-down each
contribute one half of the conductivity of the non-magnetic semiconductor. Their transport can be
regarded as occurring through separate 'spin channels', as long as the device dimensions are
shorter than the spin scattering length. When the current injected into the NMS becomes
spin-polarized, e.g. by magnetizing the DMS contacts, this implies that both spin channels in the
NMS carry a different amount of current. Because the conductivity of the spin channels is equal,
injecting a spin-polarized current - or, in other words, not using the conductivity of one of the
available spin channels - implies that the total device resistance increases. The effect can be as
large as 100\% for complete spin-polarization when only one of the two spin channels in the NMS is
used. We have performed a more detailed modeling of the device resistance using the local approach
described in Refs. \cite{gs,vanson}, and find that in a one dimensional device (Fig. \ref{fig1}a)
the resistance change is given by

\end{multicols}
\begin{equation} \label{Delta R}
 \frac {\Delta R} {R_{nms}} = \beta\frac {\lambda_{dms}}
{\sigma_{dms}} \frac {\sigma_{nms}} {x_0}
   \frac {2} {\frac {\lambda_{dms}} {\sigma_{dms}} \frac
{\sigma_{nsm}}{\lambda_{nms}}(1 + e^{-\frac{x_0}
{\lambda_{nms}}})+2\frac {\lambda_{dms}}
   {\sigma_{dms}} \frac {\sigma_{nms}}{x_0}e^{-\frac{x_0}
{\lambda_{nms}}}+1-\beta^2}
   \end{equation}
\begin{multicols}{2}

where $\lambda_{dms}, \lambda_{nms}, \sigma_{dms}, \sigma_{nms},$
are the spin flip length and the conductivity in the DMS and the
NMS respectively, $x_0$ is the spacing between the contacts and
$\beta$ is the degree of spin polarization in the bulk of the
contacts. $R_{nms}$ is given by $x_0/\sigma_{nms}$.

Eq. (1) describes a magnetoresistance effect due to spin accumulation in a non-magnetic material,
similar to the situation for GMR. However, the effect is distinct from GMR in several aspects.

The GMR effect only occurs in the limit $\lambda_{nms} > x_0$, and
corresponds to the difference in resistance between the blocking
of one spin channel at the detector (for parallel magentization of
injector and detector), and two blocked spin channels (for
antiparallel magnetization). The paramagnetic effect in this limit
results from the difference in device resistance between zero
blocked channels (for unmagnetized DMS) and one blocked channel.
In the limit of $\lambda_{nms}\gg x_0$ the maximum increase in
resistance for $\beta=1$ is indeed $x_0/\sigma_{nms}$ which is
equivalent to a doubling of the resistance of the NMS.

Much more striking perhaps is that from Eq. (1) one readily finds
that a magnetoresistance effect still exists when $\lambda_{nms}<
x_0$. In principle, also a device with only one magnetic contact
will show the effect - in contrast with GMR. In this limit, the
suppression of the spin channel occurs only over a distance of
order of the spin flip length. The behaviour of the
electrochemical potentials of the spin channels near the DMS/NMS
contact in that case is sketched in fig. \ref{fig1}c, where the
discontinuity in the average potential $\mu^\ast$ at the DMS/NMS
interface is equivalent to the boundary resistance of a
ferromagnet/nonferromagnet metal interface described by van Son et
al.\cite{vanson} and by Johnson and Silsbee\cite{johnson}. Our DMS
contacts allow for a {\it continuous tuning} of the boundary
resistance, which obviously cannot be easily realized with
ferromagnetic contacts and basically constitutes the
magnetoresistance effect in this limit - the field-induced surplus
resistance is directly related to an increase on boundary
resistance. In the experiments described below, we only employ a
geometry with two DMS contacts. This is solely because of
technological reasons, but implies that for samples where
$\lambda_{nms} < x_0$, the data simply reflect the change in
boundary resistance of two independent DMS/NMS contacts.

For an experimental demonstration of the novel magnetoresistance
effect, we have used MBE-grown II-VI-semiconductor multilayer
structures, consisting of a n-doped Zn$_{0.97}$Be$_{0.03}$Se layer
(thickness 500 nm) as a NMS, contacted by the DMS
Zn$_{0.89}$Be$_{0.05}$Mn$_{0.06}$Se (thickness 100 nm or 200 nm) ,
grown on an insulating GaAs substrate. Devices were fabricated for
a variety of doping levels above the metal-insulator transition
(which is around $n \approx 10^{18}$ cm$^{-3}$ for these
materials), i.e. aiming for nominal donor concentrations of 1, 3
and 9 $\cdot 10^{18}$ cm${-3}$ for both NMS and DMS, in all
possible combinations. The actual dopant concentrations were
determined from Hall measurements on appropriate control samples.
Contact pads ($200 \times 250  \mu$m) positioned at various
spacings ($10 \mu$m or more) were defined lithographically in a
100 nm Al layer, which was deposited on top of the semiconductor
stack to provide an ohmic contact to the DMS. These pads were then
used as an etch mask for wet chemical etching, removing the
magnetic semiconductor and some 10 nm of the
Zn$_{0.97}$Be$_{0.03}$Se in the unmasked area. In a second optical
lithography step, a mesa area including two of the DMS contact
pads was defined, and the surrounding Zn$_{0.97}$Be$_{0.03}$Se was
removed by wet chemical etching. The resulting structure is drawn
schematically in Fig. \ref{fig1}b. The magnetoresistance of a
large number of devices was measured at several different
temperatures and for fields between 0 and 7 T, using an AC voltage
bias of 100 $\mu$V (25 $\mu$V for $T<400$ mK). Care was taken to
ensure that the data were within the regime of linear response.

In the experiment, all Zn$_{0.89}$Be$_{0.05}$Mn$_{0.06}$Se /
Zn$_{0.97}$Be$_{0.03}$Se hybrid structures exhibited a strong
positive magnetoresistance. Here, we will focus on two series of
data that prove that the magnetoresistance behaviour is caused by
the effect introduced above; the data are representative for all
devices studied so far.

\begin{figure}
\centerline {\psfig{file=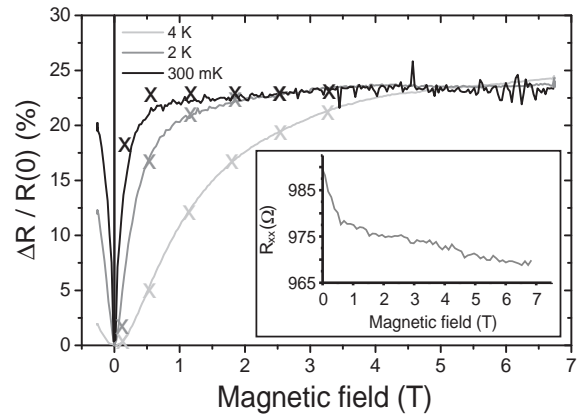,width=8.0cm}}
\caption{Relative change of total device resistance
plotted over the magnetic field for DMS/NMS multilayer structure 1671 at different temperatures
(zero field resistance is 376$\Omega$) and for a DMS Hall bar at 4.2 K (insert). The crosses
represent values obtained by fitting eq. 1 to the 4 K measurement and assuming a Boltzmann factor
for the occupation of the upper and lower Zeeman level in the DMS.} \label{fig2}
\end{figure}

Typical traces of the effect and its temperature dependence are shown in Fig. \ref{fig2} for a
device with a contact spacing of 20$\mu$m, a doping level of $n=7 \cdot 10^{18}/cm^3$ (DMS) and
$n=3.6\cdot 10^{18}/cm^3$ (NMS), and a DMS thickness of 100 nm. The maximum change in resistance is
up to 91 $\Omega$, with a total device resistance of 376 $\Omega$. A lower limit for the relative
change in resistance is $ \Delta R / R_{nms}\approx$ 25\%, which is a conservative estimate because
we neglected the contributions of contact resistances between metal and DMS (Correcting for the
contact resistance, which is known because the samples were actually fabricated in a transmission
line configuration, would yield a change of more than 30\%). We have verified that the effect does
not depend on the orientation of the magnetic field. As is also evident from fig. \ref{fig2},
reducing the temperature does not affect the saturation value, however, the saturation field is
strongly reduced - much stronger than one typically would expect from the temperature dependence of
the Zeeman splitting in the DMS. This observation reflects the strongly non-linear dependence
\cite{gs} of the spin injection efficiency on the spin-polarization in the DMS. An exact modeling
of the temperature dependence is not straight-forward because of the unknown field- and temperature
dependence of $\lambda_{dms}$ and $\lambda_{nms}$. Moreover, since Eq. (1) was derived for the
one-dimensional device of Fig. 1a and does not apply to the essentially two-dimensional devices
studied experimentally (Fig. 1b), any fits of actual data to this expression have only a limited
validity. Bearing all of this in mind, the crosses in Figure \ref{fig2} indicate the behaviour
predicted by Eq. (1), assuming $\lambda_{dms}$ and $\lambda_{nms}$ to be temperature independent,
and fitted to the magnetoresistance behaviour at 4 K. The crosses for the other temperatures were
obtained simply by assuming a Boltzmann distribution of the conduction electrons between the Zeeman
levels, i.e., neglecting band-filling effects\cite{bandfilling} while keeping the other parameters
constant. Evidently, Eq. (1) gives a reasonable description of the actually observed device
behaviour. However, we should note that below $\approx$ 0.3 T the fit to the experiment is less
accurate.

Actual values for the free parameters in Eq.(1) can be obtained
from the saturation magnetoresistance using $\sigma_{dms} \approx
2 \cdot 10^{2}$ $\Omega^{-1}$cm$^{-1}$ and $\sigma_{nms} \approx
1.5 \cdot 10^{2}$ $\Omega^{-1}$cm${-1}$, as obtained from control
samples, yielding $\lambda_{dms} \approx 20$ nm and $\lambda_{nms}
\approx 1.5 \mu$m. The spin polarization $\beta$ in the DMS is
deduced from the Zeeman splitting as obtained from optical
experiments. The values for the spin scattering length obtained
from this fit seem quite reasonable; $\lambda_{dms}$ is of a
similar magnitude as the values usually encountered for
ferromagnetic metals, and $\lambda_{nms}$ agrees well with optical
data by Kikkawa et al.\cite{Kik}.

\begin{figure}
\centerline {\psfig{file=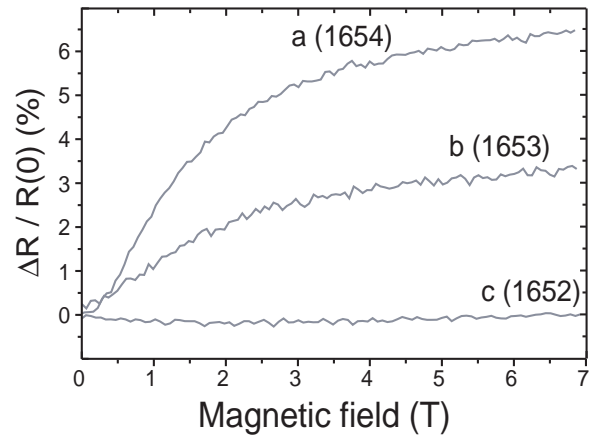,width=8.0cm}} \caption{Relative change of total device resistance
plotted over the magnetic field for three different devices. 1653 (b) and 1654 (a) are spin
injection devices with DMS thickness of 100 and 200 nm, respectively. The zero field resistance is
15 $\Omega$ (1653) and 20.5 $\Omega$ (1654). The relative magnetoresistance increases by a factor
of 2 when the DMS thickness is doubled. 1652 (c) is a reference device without DMS ($R_0$=14
$\Omega$). The magnetoresistance is less than 0.02 $\Omega$.} \label{fig3}
\end{figure}

Fig. \ref{fig3} displays experimental results obtained for a
series of devices with doping levels
$n_{nms}=8.6\cdot10^{18}/cm^3$, $n_{dms}=4\cdot10^{18}/cm^3$,
leading to conductivities $\sigma_{nms} \approx 3 \cdot 10^{2}$
$\Omega^{-1}$cm$^{-1}$ and $\sigma_{dms} \approx 1 \cdot 10^{2}$
$\Omega^{-1}$cm$^{-1}$, a contact spacing of 10$\mu$m, and a DMS
thickness of 0, 100 and 200 nm. From these data, three major
features are apparent. First, the maximum size of the relative
magnetoresistance effect is reduced to about 6\%. This can be
explained by the increased spin scattering in the higher doped
NMS.\cite{Kik} Second, reducing the DMS thickness from 200 (fig.
\ref{fig3}a) to 100 nm (fig. \ref{fig3}b) results in a reduction
of the relative effect by a factor of 2. This observation can be
understood by realizing that, due to the finite spin scattering
length in the DMS, a thinner DMS layer results in a lower degree
of spin polarization, again showing that the effect is quite
sensitive to even a small number of electrons in the upper Zeeman
level. Finally, curve (c) was measured for a reference device
where the DMS layer was omitted. In this case, only a small
($<1\%$) negative magnetoresistance is observed, possibly due to
weak localization effects. This observation clearly evidences that
spin injection via the DMS layer is an absolute necessity to
observe the novel magnetoresistance effect. Using Eq. (1) and the
conductivities quoted above, we can consistently reproduce the
observed DMS thickness-dependence of the effect for $\lambda_{dms}
\approx 35$ nm and $\lambda_{nms} \approx 0.5 \mu$m, which again
is in line with the observed\cite{Kik} decrease of spin scattering
length with increasing dopant concentration in highly doped
samples.

The high dopant concentration in the DMS layers in the devices of
Figs. 2 and 3 was chosen to guarantee that the intrinsic
magnetoresistance of the DMS is negative. At lower n (but above
the metal-insulator transition), DMSs show an additional positive
magnetoresistance due to the e-e  correction to the conductivity.
This correction vanishes with increasing $n$, according to
$(k_{\rm F}l)^{-3/2}$\cite{dietl}, where $k_{\rm F}$ is the
wavevector at the Fermi energy and $l$ is the mean free path of
the electron. That we are indeed in the limit where only the weak
localization correction to the conductivity remains is evidenced
by the small negative magnetoresistance (2\%) of a sample
consisting only of DMS (Fig. \ref{fig2}, insert).

In order to better understand the device behavior and the magnitude of the effect, we have
performed two dimensional simulations of the current flow in the device based on the
drift-diffusion equation and incorporating the model of reference\cite{gs}. The details of these
calculations will be presented elsewhere, but here we will summarize the main findings relevant to
the present paper.

(i) Because of the relatively low conductivity in the DMS layer,
the current into the NMS layer is injected perpendicular to the
DMS/NMS interface across the whole width (200$\mu m$) of the
contact pad. As already indicated above, this raises questions on
the validity of using Eq. (1) for extracting material parameters
from the actual measurements.

(ii) Because of this current profile, the resistivity of the DMS
layer contributes only on the order of 2\% to the total device
resistance. This implies that the intrinsic magnetoresistance of
the DMS (Fig. \ref{fig2}-insert) can be neglected when describing
the overall device magnetoresistance.

(iii) The total device resistance is mainly determined by the
region under the contact which is close to the DMS/NMS interface.

(i) and (iii) together imply that one can have sizeable
magnetoresistance effects, even when the spin-scattering length is
in the  sub-micron regime, but cannot expect to see a dependence
of the effect on the spacing between the contacts. In order to
observe the latter, one needs to fabricate sub-micron contacts on
a micrometer sized mesa. However, such a technology does not yet
exist.

Given the small and negative magnetoresistance measured for the
reference layers (Fig. \ref{fig2}-insert, \ref{fig3}), we conclude
that our experimental data on the multilayer samples directly
evidence that we have succeeded in observing the novel,
spin-injection-induced, magnetoresistance effect described above.
Note that in contrast to Ref. \cite{injector} the data presented
here were all taken in the regime of linear response. They
represent a very strong evidence for the single-particle character
of the electrical spin injection from a DMS and confirm the
validity of Ref. \cite{gs,vanson,johnson} in describing the
injection phenomena. The strong temperature dependence of the
saturation behavior in a regime where the giant Zeeman splitting
is almost temperature-independent is further evidence that the
polarization in a diffusive spininjector has to be very close to
unity in order to achieve efficient spin injection. At the same
time, our data demonstrate a new magnetoresistance effect which
can be regarded as the paramagnetic version of GMR. The results
illustrate a viable route towards a straightforward determination
of spin polarization in semiconductor devices. Using spin
dependent resistance effects spin controlled programmable logic
may become feasible; other applications could be found in read-in
and read-out mechanisms for solid state quantum-computing with
spins.

\acknowledgments

We acknowledge the financial support of the Bundesministerium
f\"ur Bildung und Forschung and the European Commission. We would
also like to thank V. Hock, N. Schwarz, T. Dietl, and G. E. W.
Bauer for help and discussions.

\end{multicols}


\begin{references}
\bibitem{1d} For technological reasons, an exact realization of the
one-dimensional device in fig. 1a is not feasible;
the main problem is the high contact resistance between Al and the
II-VI compounds.
\bibitem{bandfilling} This approximation is not straightforward.
When one assumes a parabolic conduction band in the DMS, the
doping levels used here would imply a Fermi energy of order tens
of meV, comparable to or larger than the Zeeman splitting of the
conduction band, and leading to reduced spin polarization. This is
in contradiction with our experimental results on the spin diode
(Ref. 1). However, since the DMSs are doped much above the Mott
criterium, the parabolic band model is not applicable. The
metallic side of the Mott transition directly implies the
existence of an impurity band, which has a much higher density of
states than one would estimate from a parabolic band model.
Temperature-dependent transport measurements by our group support
the existence of an impurity band.
\end{references}
\end{document}